\documentclass[aps,prl,twocolumn,superscriptaddress,showpacs,floatfix]{revtex4-1}
\usepackage{amsmath,amssymb,amsfonts,dsfont, float,graphics,epsfig,epstopdf,color,verbatim,tabularx,bm,multirow}
\usepackage{wasysym}
\usepackage[utf8]{inputenc}
\usepackage[T1]{fontenc}
\usepackage{xcolor}
\usepackage{ulem}
\usepackage{soul}
\usepackage{cancel}

\IfFileExists{newtxtext.sty}
   {\usepackage{newtxtext,newtxmath}}
   {\IfFileExists{stix.sty}
      {\usepackage{stix}}
      {\IfFileExists{mathptmx.sty}
      {\usepackage{mathptmx}}{} } }

\usepackage{hyperref}
\hypersetup{
	colorlinks = true,
	linkcolor = [rgb]{0.70,0.13,0.13},
	citecolor = [rgb]{0.13,0.55,0.13},
	urlcolor = [rgb]{0.25,0.41,0.88}}

\DeclareSymbolFont{sfletters}{OML}{cmbrm}{m}{it}
\DeclareMathSymbol{\sfeps}{\mathord}{sfletters}{"22}

\graphicspath{{Figures/}}

\begin{document}
\title{Entanglement entropy and disorder operator at kagome deconfined quantum criticality}

\author{Yan-Cheng Wang}
\email{ycwangphys@buaa.edu.cn}
\affiliation{Hangzhou International Innovation Institute, Beihang University, Hangzhou 311115, China}
\affiliation{Tianmushan Laboratory, Hangzhou 311115, China}

\author{Zheng Yan}
\email{zhengyan@westlake.edu.cn}
\affiliation{Department of Physics, School of Science and Research Center for Industries of the Future, Westlake University, Hangzhou 310030,  China}
\affiliation{Institute of Natural Sciences, Westlake Institute for Advanced Study, Hangzhou 310024, China}

\author{Xue-Feng Zhang}
\email{zhangxf@cqu.edu.cn}
\affiliation{Department of Physics, Chongqing University, Chongqing 401331, China}

\begin{abstract}
We investigate the deconfined quantum critical point (DQCP) candidate in the extended hard-core Bose-Hubbard model on the kagome lattice, employing quantum Monte Carlo simulations to study the entanglement entropy and the $U(1)$ disorder operator. In stark contrast to findings in $J$-$Q$ models and other candidates, the universal logarithmic correction coefficients for both quantities are found to be {positive}, consistent with a unitary conformal field theory (CFT). Crucially, the current central charge $C_J$, extracted from the small-angle behavior of the disorder operator, is enhanced by a factor of approximately {4/3} compared to that of the conventional 3D $O(2)$ Wilson-Fisher fixed point. This enhancement {implies} a consistent explanation in the recently observed low-energy excitation spectrum at this DQCP, which features {two distinct linearly dispersing modes} with a velocity ratio of approximately three. Our results provide evidence that this quantum phase transition constitutes a genuine DQCP, characterized by coexisting fractionalized excitations that collectively modify its critical properties.

\end{abstract}
\date{\today}
\maketitle

{\color{blue}\it{Introduction.-}} 
The theory of deconfined quantum criticality, proposing direct continuous quantum phase transitions outside the Landau-Ginzburg-Wilson symmetry-breaking paradigm, has fundamentally reshaped our understanding of quantum matter~\cite{senthilDeconfined2004, senthilQuantum2004}. Such deconfined quantum critical points (DQCPs) are predicted to host a tapestry of exotic phenomena: emergent gauge fields, fractionalized (deconfined) excitations, and enhanced global symmetries~\cite{senthilfisher, nahumEmergent2015, maRole2019}. This framework offers a compelling theoretical lens for interpreting transitions between ordered states with incompatible symmetry breakings, such as between a N\'eel antiferromagnet and a valence bond solid (VBS)~\cite{sandvikEvidence2007, nahumDeconfined2015, qinDuality2017, wangDeconfined2017}, or between a superfluid and a VBS~\cite{kagome_Senthil,kagome_wessel,kagome_zhang,Zhang2018PRL}. Beyond their theoretical intrigue, DQCPs hold significant potential for understanding the exotic properties of quantum materials and for informing designs in quantum simulation~\cite{jimenezquantum2021, cuiProximate2023, guoDeconfined2025}.

However, the concrete realization of a stable, conformal DQCP in microscopic lattice models has become a central and contentious puzzle in modern quantum many-body physics. Intensive numerical scrutiny of the seminal $J$-$Q$ model on the square lattice---long considered the archetypal DQCP candidate---has revealed persistent complications. Accumulating evidence, including critical exponents violating conformal bootstrap bounds~\cite{nahumDeconfined2015, nakayamaNecessary2016}, anomalous negative logarithmic corrections in entanglement entropy scaling~\cite{JRZhao2021DQC}, and detailed finite-size scaling analyses~\cite{deng2024diagnosing}, increasingly suggests that the transition may be weakly first-order or described by a non-unitary conformal field theory (CFT) \cite{RCMa2020}. Recent high-precision studies using the fuzzy sphere method have further solidified the view that the presumed $SO(5)$ symmetric DQCP in this setting exhibits first-order characteristics~\cite{zhou2024mathrmso5}. These findings have ignited a vigorous debate about the very existence of a stable DQCP described by a unitary CFT in local, bosonic lattice Hamiltonians. Similar ambiguities between continuous and weakly first-order transitions extend to other proposed settings, underscoring the challenge~\cite{kuklovDeconfined2008, jiangFrom2008, chenDeconfined2013}.

Amidst this searching landscape, the extended hard-core Bose-Hubbard model (EHBHM) on the kagome lattice stands out as a particularly promising and distinct candidate. This model exhibits a direct, numerically continuous quantum phase transition at 1/3 filling between a $U(1)$-symmetry-breaking superfluid and a {translational}-symmetry-breaking VBS phase---symmetries that are incompatible under a Landau description~\cite{Zhang2018PRL}. Early studies have documented key signatures consistent with DQC, such as a large anomalous dimension for the VBS order parameter. Most intriguingly, and in stark contrast to other candidates, recent high-resolution spectral functions have revealed a defining dynamical fingerprint: the presence of \textit{two distinct low-energy, linearly dispersing excitations} with a large velocity ratio of approximately three~\cite{liu2024deconfined}. This clear splitting fundamentally violates Lorentz invariance at the critical point and provides strong evidence for the coexistence of multiple fundamental excitations---a hallmark of deconfinement and fractionalization. This unique spectral feature makes the kagome lattice system a critical testing ground, demanding a comprehensive investigation through the lens of universal, non-local observables to determine the true nature of its criticality.

\begin{figure*}[htp!]
	\centering
	\includegraphics[width=0.9\textwidth]{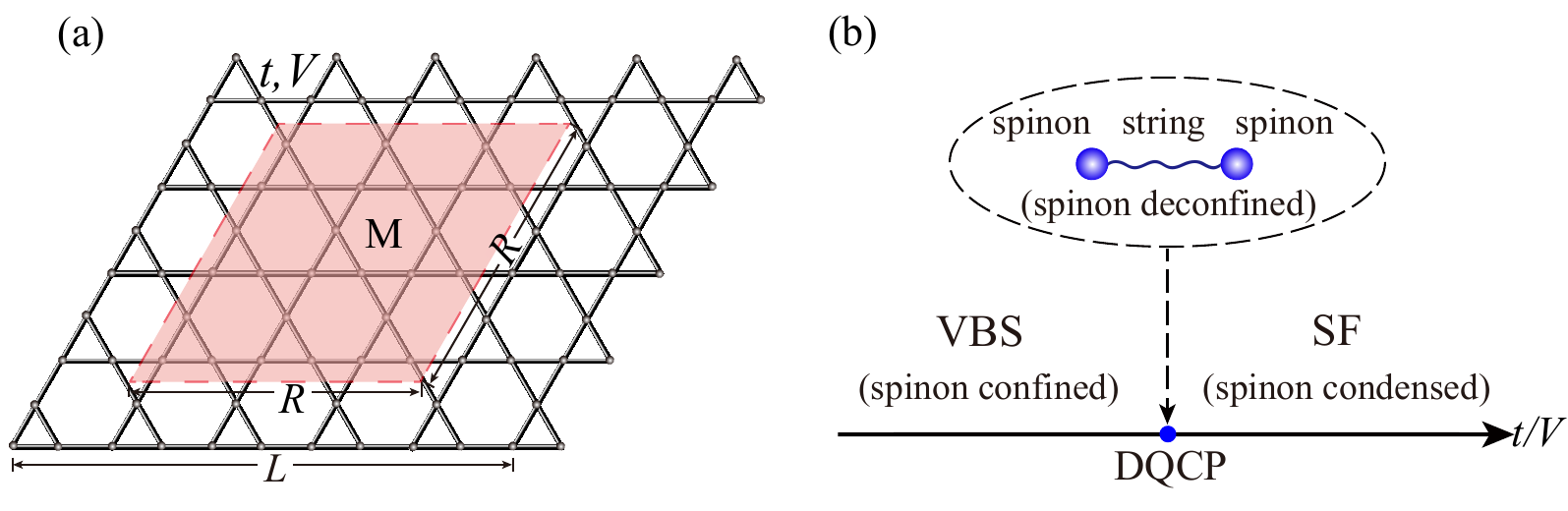}
	\caption{(a)Kagome lattice of {extended} hard-core Boson-Hubbard model: disorder operator $X_M$ and entanglement entropy $S^{(2)}$ are applied on regions $M$ with size $R \times R$ and perimeter $l=4R-4$ in the $L \times L$ lattice. (b) The schematic {quantum} phase diagram of the {EHBHM} at $1/3$ filling.}
	\label{fig:fig1}
\end{figure*}

In this work, we perform such a definitive investigation by combining high-precision quantum Monte Carlo {(QMC)} simulations of universal scaling properties with insights from the recent study of dynamical spectrum \cite{liu2024deconfined}. We focus on two fundamental non-local probes: the second R\'enyi entanglement entropy and the $U(1)$ disorder operator. We find that, unlike in the $J$-$Q$ model, both quantities exhibit \textit{positive universal logarithmic corner corrections} at the critical point, a signature fully consistent with a unitary CFT and opposing the trend seen in weakly first-order scenarios. This already marks a significant departure from the prevailing narrative. Our central and most revealing result comes from a detailed analysis of the disorder operator, which allows us to cleanly extract the current central charge, $C_J$. We find that $C_J^{\text{DQCP}}$ is enhanced by a factor of $1.33(5)$ compared to the established value for the conventional 3D $O(2)$ Wilson-Fisher fixed point.

We then synthesize these static and dynamical observations into a unified physical picture. We argue that the observed enhancement of $C_J$ is not merely coincidental but is the direct, quantitative consequence of the two-mode low-energy structure. By positing that both the slower mode (associated with {quantum} string {\cite{Rydberg_string1,Rydberg_string2}}) and the faster mode (associated with deconfined spinons) contribute comparably to the current correlation function, and noting that such contributions are inversely proportional to the excitation velocities, we derive the simple relation $C_J \propto \sum_i g_i/v_i$. Using the experimentally measured velocity ratio $v_f/v_s \approx 3$, this model naturally predicts an enhancement ratio of $C_J^{\text{DQCP}}/C_J^{O(2)} \approx 4/3 \approx 1.33$, in remarkable agreement with our direct numerical measurement. This work thus achieves a rare synthesis: it not only provides robust evidence from scaling laws that the kagome lattice transition is a continuous DQCP described by a unitary CFT but also pinpoints the microscopic origin of its distinguishing universal coefficient in the unique, velocity-split excitation spectrum characteristic of deconfinement.


{\color{blue}\it{Model.-}}
The {EHBHM} on the kagome lattice, as shown in Fig~\ref{fig:fig1}, {and} its Hamiltonian takes the following form,
\begin{equation}
\centering
H=-t\sum_{\langle i,j \rangle}b_i^\dagger b_j + h.c. + V\sum_{\langle i,j \rangle}n_i n_j - \mu \sum_{i} n_i,
\label{eq:eq1}
\end{equation}
where $b^\dagger (b)$ is the boson creation (annihilation) operator, $n_i =b^\dagger b$ is the boson number, $t > 0$ is the hopping between nearest-neighbor sites on the kagome lattice, $V > 0$ is the nearest-neighbor repulsion, and $\mu$ is the chemical potential. {The hard-core boson can be exactly mapped to the spin 1/2, so the Hamiltonian is also equivalent to the XXZ model in a longitudinal field.} The ground state of this model has two phases: a valence bond solid phase for large $V/t$ which breaks the {translational} symmetry of the lattice and a superfluid (SF) phase for small $V/t$, which breaks the $U(1)$ symmetry of spins in the x-y plane, separated by a phase transition. In an exact $1/3$ filling, the transition is thought of as a {DQC} since the broken symmetries are not compatible under the Landau-Ginzburg-Wilson theory, but the numerical results are continuous at the quantum phase transition point~\cite{Zhang2018PRL}. At this filling, the critical point is located at $t_c/V=0.130262(3)$, $\mu/V=0.8735(1)$, determined from the previous work~\cite{Zhang2018PRL}.

{\color{blue}\it{Entanglement entropy.-}} To further explore the candidate of DQCP in this {EHBHM}, we employ the non-equilibrium increment method within the framework of {QMC} simulations~\cite{Syljuaasen2002,Sandvik2010,ZYMeng2008,yan2019sweeping,yan2020improved,albaOut2017,demidioEntanglement2020,zhaoScaling2022} to detect the second R\'enyi entropy of the kagome {DQCP}. The second R\'enyi entanglement entropy is defined as $S_M^{(2)}=-\ln[ \mathrm{tr} (\rho_M^2)]$, where $M$ is the entangled subsystem as shown in Fig.\ref{fig:fig1} {(a)} and $\rho_M$ is its reduced density matrix of ground state. We calculate EE $S_M^{(2}$ of region $M$ with $R=L/2$ for each system size $L$ and $S_M^{(2}(L)$ has the following scaling:
\begin{equation}
\centering
	S_M^{(2)}(L)=a L-s \ln L + b.
\label{eq:eq2}
\end{equation}

\begin{figure}[htp!]
	\centering
	\includegraphics[width=\columnwidth]{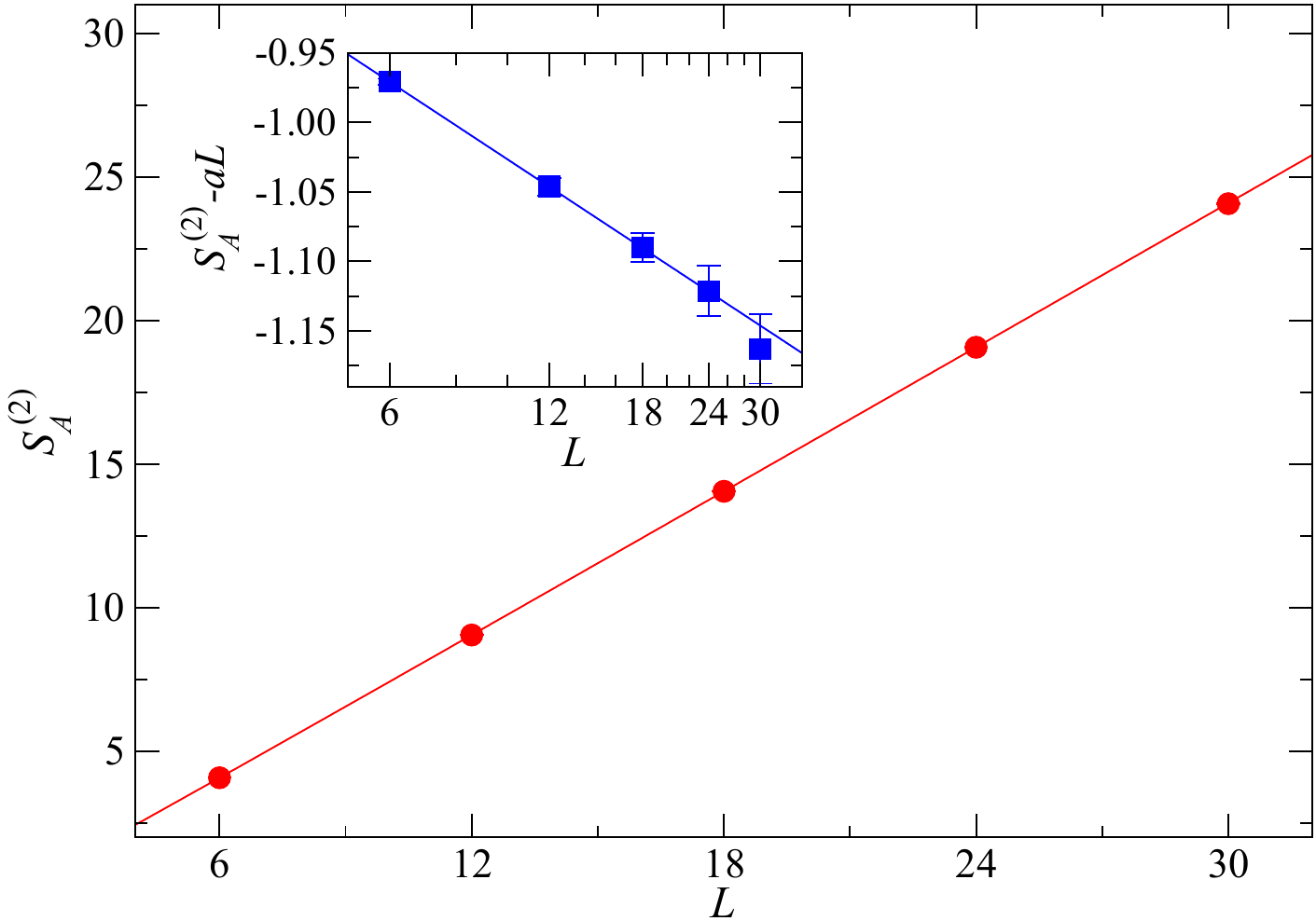}
	\caption{The second order R\'enyi entanglement entropy  $S_M^{(2)}$ as function of $l$ at the {DQCP}  ($t_c/V=0.130262$) with $\beta=50 L/3$. The fitting result is $S_M^{(2)}(L)=0.8413(3)L-0.108(3)\ln{L}-0.776(3)$. Inset shows the $S_M^{(2)}(L)-aL$ versus $L$ such that the sign of the log-corrections manifest.} 
	\label{fig:fig3}
\end{figure}

The fitting results of the Eq.\ref{eq:eq2} are shown in the Fig. \ref{fig:fig3}, the subleading coefficient $s$ of the correction term $\ln L$ is a positive number $\sim 0.1$ which is consistent with a unitary CFT prediction, and the value is close to the result predicted by a $O(N)$ Wilson-Fisher phase transition~\cite{JRZhao2021DQC,YCWang2021,wang2024quantummontecarloalgorithm} (this value almost unchanges by $N$).
It appears to be a continuous phase transition and is totally different from the result in the $J-Q$ model~\cite{JRZhao2021DQC,song2024extracting,deng2024diagnosing} which is actually a weakly first-order phase transition with an emergent $SO(5)$ symmetry breaking \cite{deng2024diagnosing}. 

{Therefore, as far as} we know, this is the only candidate for DQCP with a negative correction of the $\ln L$ term ($s>0$) on the EE scaling so far. However, since the corners of the entangled region in the kagome lattice cannot be right angles ($\pi/2$), the contribution from corner-sharps is theoretically difficult to determine. In contrast, the contribution of non-right angles can be well-defined in the disorder operator, so we turn to calculate the scaling behaviors of the disorder operator at the DQCP here to extract more information.

{\color{blue}\it{Disorder Operator.-}}
We now focus on the disorder operator in a system with $U(1)$ symmetry. Given a region $M$ on the lattice, as illustrated in Fig.\ref{fig:fig1}, we can define the disorder operator $X_M(\theta)=\prod_M e^{i\theta(S^{z}_{i}-\frac{1}{2})}$, where $S^{z}_{i}-\frac{1}{2}$ is the boson number on site $i$ since the system has a $U(1)$ symmetry in the $x$-$y$ plane. The ground state expectation \textit{\(|\langle X(\theta) \rangle|\)}  is the module of  $\langle X(\theta) \rangle$ defined as the disorder parameter that can extract the order and high-form symmetry of the disorder phase. The scaling behaviors of $X(\theta)$ rely on whether the phase is ordered or disordered. In the disordered phase, $|\langle X(\theta) \rangle|$ is proportional to $e^{-a(\theta)l}$, where $l$ is the perimeter of the region $M$. In the $U(1)$ symmetry breaking ordered phase, it usually satisfies $|\langle X(\theta) \rangle| \sim e^{-b(\theta)l\ln l}$ ~\cite{YCWang2021,Wu2021,Estienne_2022,lake2018higherform}. More interestingly, the logarithmic correction term will appear in the scaling behavior of $|\langle X(\theta) \rangle|$ at quantum critical point (QCP). The previous analytical and numerical works pointed out that $|\langle X(\theta) \rangle|$ can hold the following form at (2+1)d QCP~\cite{YCWang2021,XCWu2020,wangScaling2021,wangScaling2022,liu2024measuring}:
\begin{equation}
\centering
	\ln |\langle X_M(\theta)\rangle|=-a_1 l+s \ln l + a_0.
\label{eq:eq3}
\end{equation}
Here, the dependence on $\theta$ for all the coefficients is suppressed.
The logarithmic correction, which translates into a power law $l^s$ in $|\langle X_M\rangle|$, originates from sharp corners of the region. In general, $s$ is a universal function of both $\theta$ and the angles of the corners. 
The similar contributions from the corner were known to exist for R\'enyi entropy in a CFT, which can be understood as the disorder parameter of the replica symmetry. 
The corner correction is thought as a generic feature for disorder operators in any $(2+1)d$ CFT~\cite{YCWang2021,wangScaling2022,LiuF2023,liu2023disorder}. 

\begin{figure*}[htp!]
	\centering
	\includegraphics[width=\linewidth]{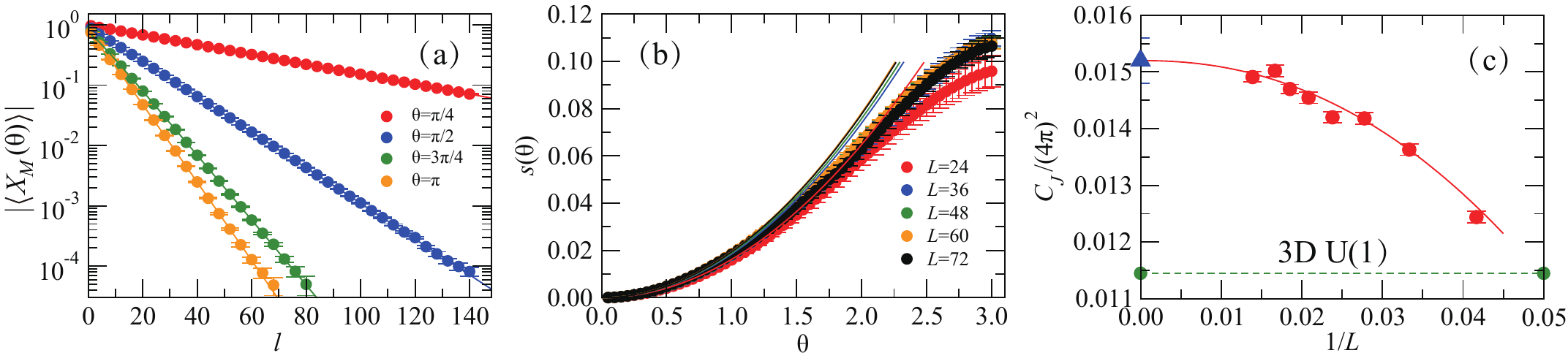}
	\caption{(a) Disorder parameter $|\langle X_M(\theta) \rangle|$ as function of $l$ at the {DQCP} ($t_c/V=0.130262$) with $\theta=\pi/4, \pi/2, 3\pi/4,\pi$ and system sizes up to $L=72$, $R\in [1,L/2]$. Dots are the QMC results with error bars smaller than the symbol size and the solid lines show the fit by the function in Eq.~\eqref{eq:eq3}. (b) The $\theta$ dependence of universal coefficient $s$ for different system sizes. For small $\theta$, a quadratic dependence clearly manifests. (c) The finite-size results converge for $L$ and fitting with Eq.~\eqref{eq:eq6} yields the coefficient $\frac{s}{f_M\theta^2}\approx 0.0152(5)$, about {1/3} times larger than that of the 3D U(1) where ${C_J}/{(4\pi)^2}=0.01145$.}
	\label{fig:fig2}
\end{figure*}

We choose the region $M$ to be a diamond region in the lattice with perimeter $l=4R-4$ as Fig.~\ref{fig:fig1} shows, which has two $\pi/3$  corners and two $2\pi/3$ corners. To calculate the disorder operator of the EHBHM, we employ large-scale stochastic series expansion (SSE) QMC simulations~\cite{Syljuaasen2002,Sandvik2010,ZYMeng2008,yan2019sweeping,yan2020improved}, and compute the expectation value of $X_M(\theta)$ on a finite lattice with $\beta=1/T=25L/3$ and $R\in[1,L/2]$ to access the thermodynamic limit.

According to the Taylor expansion on a small $\theta$ of $|\langle X_M\rangle|$, the subleading coefficient satisfies:
\begin{equation}
\centering
	s(\theta)=f_M \frac{C_J}{(4\pi)^2} \theta^{2},
\label{eq:eq6}
\end{equation}
where $C_J$ is the current central charge depending on the universality in the CFT and the factor $f_M$ is contributed by the geometric of corners.
In our case with two $\pi/3$  corners and two $2\pi/3$ corners, 
	$f_M=\frac{2f(\pi/3)+2f(2\pi/3)}{4f(\pi/2)}\approx1.3023$~\cite{LiuF2023},
where the $f(\alpha)$ is decided by the angle $\alpha $ in the region $M$:
	$f(\alpha)=2(1+(\pi-\alpha)\cot(\alpha))$~\cite{Wu2021,liu2023disorder}.

We now turn to the critical point $t_c/V=0.130262$. 
{Fig.~\ref{fig:fig2}(a) displays numerical results of $\langle X_M(\theta)\rangle$ versus $l$ for several $\theta$ values. The data are well described by the scaling form in Eq.~\eqref{eq:eq3}. In particular, the subleading logarithmic coefficient $s$, plotted in Fig.~\ref{fig:fig2}(b), exhibits a clear quadratic dependence on $\theta$ for small $\theta$.}
To corroborate the analytical results, we examine more closely the function $s(\theta)$ as $\theta\rightarrow 0$. The coefficient $s(\theta)$ exhibits a clear $\theta^2$ dependence, and it is found to be $0.0152(5)$ for the direct fitting method, as shown in Fig.~\ref{fig:fig2}(c). Therefore, we find the {central charge} of this DQCP is larger than the one of 3D U(1) criticality:
\begin{equation}
\centering
	\frac{C_J^\mathrm{DQCP}}{C_J^{3\mathrm{D}\  U(1)}}\approx1.33(5)
\label{eq:eq6}
\end{equation}
where the ${C_J^{3\mathrm{D}\  U(1)}}=1.8088$ has been calculated from the $O(2)$ Wilson-Fisher CFT~\cite{Katz2014,Krempa2014,Chester_2020}. Actually, the {central charge} of normal Wilson-Fisher fixed points, such as $O(3)$ and $O(4)$ CFT, are almost same as $O(2)$ according to the conformal bootstrap~\cite{Poland2019}. 

It should be emphasized that the scaling behavior of the disorder operator presented here differs fundamentally from what is observed in the $J-Q$ models~\cite{wangScaling2022}. The positive coefficient of $\ln l$ term at large $U(1)$ rotation ($\theta \to \pi$), strongly supports this may be a true DQCP, among the realistic models so far as we know. In the following, we will further analyze the reason for the large current {central charge} here.

{\it{\color{blue} Current central charge and excitations.-}}
We find that the large current {central charge} $C_J^{DQCP}$ can be understood via the recent study on the spectroscopy at the DQCP which pointed out that there are two CFTs coexisting with different velocities~\cite{liu2024deconfined}. The low-energy {excitation spectrum} at this DQCP solved by QMC features {two distinct, linearly dispersing modes}. Their characteristic propagation velocities differ significantly, with a ratio of roughly $v_2 / v_1 \approx 3$.
We think that these two facts (current {central charge} and velocities) are not independent but are causally linked through the fundamental structure of the DQCP low-energy theory.

In a CFT, the current central charge $C_J$ quantifies the strength of the two-point correlation function for a conserved current $J^\mu$:
\begin{equation}
\langle J^\mu(x) J^\nu(0) \rangle = \frac{C_J}{(x^2)^{\Delta_J}} I^{\mu\nu}(x),
\end{equation}
where $\Delta_J = d-1$ is the protected scaling dimension of the conserved current and $I^{\mu\nu}(x)$ is a fixed tensor structure~\cite{francesco2012conformal}. While $C_J$ is often interpreted as a measure of the number of degrees of freedom charged under the corresponding symmetry, its value in an interacting low-energy effective theory is more precisely a {weighted sum over contributions from all gapless excitations} that couple to $J^\mu$.


At the DQCP, the spectrum contains two low-energy branches with velocities $v$ and $3v$. We note that the $v$ mode is caused by the quantum string fluctuations and located at the {$K$} momentum point, and the $3v$ mode is introduced by the deconfined spinons and located at the {$\Gamma$} momentum point~\cite{liu2024deconfined,zhou2020string}. 
The striking numerical agreement between the enhanced $C_J$ and the simple model based on the two-mode spectrum suggests a deep connection. A natural, albeit phenomenological, hypothesis is that both gapless modes contribute to the current correlation function. For example, when the spinon (fast mode) moves into (or out of) the region $M$, the kink (slow mode) on the attached quantum string can also cross the boundary. Given the observed velocities $v_K = v$ and $v_{\Gamma} = 3v$, a simple conjecture of their combined contribution is:
\begin{equation}
C_J^{\text{DQCP}} \approx \left(1+\frac{v_K}{v_\Gamma}\right){C_J^{U(1)}} \approx \frac{4}{3} {C_J^{U(1)}},
\end{equation}
{which} yields a direct prediction:
$\frac{C_J^{\text{DQCP}}}{C_J^{U(1)}} \approx 4/3$.

Therefore, the increase in $C_J$ is not merely correlated with but is causally explained by the specific structure of the low-energy excitation spectrum at the DQCP. The enhancement quantifies the additional contribution from the strong interplay between the quantum string and deconfined fractionalized spinon, but nonetheless, leads to a significant net increase over the conventional critical point. This synthesis of static and dynamical critical data presents robust evidence for the deconfined nature of the quantum critical point.

{\color{blue}\it{Conclusion.-}}
We have presented a comprehensive numerical study of the scaling of entanglement and disorder operator at the QCP between superfluid and VBS phases in the kagome lattice EHBHM. The positive logarithmic correction coefficients in both the entanglement entropy and the $U(1)$ disorder operator starkly contrast with the negative sign observed in other prominent DQCP candidates such as the $J$-$Q$ model, suggesting this phase transition is a critical point rather than a weakly first-order transition.

The most salient finding is the significant enhancement of the current central charge, $C_J^{\text{DQCP}} / C_J^{U(1)} \approx 4/3$. We have conjectured a physical mechanism for this enhancement, linking it directly to the unique low-energy excitation spectrum recently revealed at this DQCP \cite{liu2024deconfined}. The presence of two gapless modes---a slower mode associated with quantum fluctuations of the string (domain-wall) and a faster mode from deconfined spinons---with velocities differing by a factor of $\sim 3$, naturally explains the observed $C_J$ enhancement with a factor $4/3$. 

Together, the positive universal logarithmic terms in non-local observables and the enhanced current central charge explained by a two-mode fractionalized spectrum offer a consistent and hopeful evidence. This supports the scenario that the quantum phase transition in this kagome lattice model realizes a true DQCP, where emergent fractionalization with the generalized higher-form symmetries and unconventional critical dynamics are key defining features. Our work not only identifies a promising candidate for stable DQCP physics but also demonstrates the powerful synergy between entanglement/disorder diagnostics and dynamical spectral information in probing the essence of exotic quantum criticality.

{\color{blue}\it{Acknowledgement.-}}
We thank Meng Cheng, Zi Yang Meng, Hong-Hao Tu, Zheng Zhou, and Weilun Jiang for valuable discussions on the related topic.
YCW acknowledges the support from the Natural Science Foundation of China (Grant No. 12474216) and Zhejiang Provincial Natural Science Foundation of China (Grant No. LZ23A040003), and the Start-up Funding of Hangzhou International Innovation Institute of Beihang University.
ZY is supported by the Scientific Research Project (No. WU2025B011) and the Start-up Funding of Westlake University. 
X.-F. Z. acknowledges funding from the National Science Foundation of China under Grants  No.12274046 and No.12547101, and Xiaomi Foundation / Xiaomi Young Talents Program.
The authors thank the high-performance computing centers of Hangzhou International Innovation Institute of Beihang University and of Westlake University and the Beijng PARATERA Tech Co.,Ltd. for providing HPC resources.


\bibliographystyle{apsrev4-1}
\bibliography{kgdqcp}

\setcounter{equation}{0}
\renewcommand{\theequation}{S\arabic{equation}}
\renewcommand{\thefigure}{S\arabic{figure}}

\newpage

\begin{widetext}
	

\end{widetext}

\end{document}